\documentclass[10pt, a4paper, floats,booktabs,amsmath,amssymb]{article}

\usepackage{multicol}

\usepackage{graphicx,amsmath}
\usepackage{bm}
\usepackage{subcaption}
\usepackage{subfiles}
\usepackage{enumerate}
\usepackage[inline]{enumitem}
\usepackage{multirow}
\usepackage{array}
\usepackage{booktabs}
\usepackage{siunitx}

\usepackage{blkarray}
\usepackage{mathtools}
\usepackage{stfloats} 
\usepackage{tikz}
\usepackage{adjustbox}
\usetikzlibrary{quantikz2}
\usepackage{adjustbox}
\usepackage{float}
\usepackage{nicematrix}
\usepackage{algorithmic}
\usepackage{array}
\usepackage{lipsum}
\usepackage{xcolor}
\usepackage{hyperref}

\hypersetup{
    colorlinks,
    linkcolor={red!80!black},
    citecolor={blue!50!black},
    urlcolor={blue!80!black}
}

\newcommand\RED{\color{red}}
\newcommand{\rdots}{\hspace{.2ex}\raisebox{1ex}{\rotatebox{-12}{$\ddots$}}}

\usepackage{diagbox}
\usepackage{bibentry}
\usepackage{natbib,hyperref}

\def\thickhline{\noalign{\hrule height.9pt}}

\newcommand\Rey{\mbox{\textit{Re}}}   
\newcommand\Mach{\mbox{\textit{Ma}}}  
\newcommand{\ignore}[1]{}
\newcommand{\nobibentry}[1]{{\let\nocite\ignore\bibentry{#1}}}

\begin{document}


\title{
Quantum Unitary Matrix Representation of Lattice Boltzmann Model for Low Reynolds Fluid Flow Simulation
}

\author{E. Dinesh Kumar$^*$ and Steven H. Frankel$^\dagger$ \vspace{0.3em} \\
\textit{Faculty of Mechanical Engineering} \\
\textit{Technion - Israel Institute of Technology, Haifa, 3200003, Israel.}\\
{$^*$edk261@gmail.com and $^\dagger$frankel@technion.ac.il}
} 
%



\maketitle


\begin{abstract}

We propose a quantum algorithm for the Lattice Boltzmann (LB) method to simulate fluid flows in the low Reynolds number regime. 
First, we encode the particle distribution functions (PDFs) as probability amplitudes of the quantum state and demonstrate the need to control the state of the ancilla qubit during the initial state preparation. Second, we express the LB algorithm as a matrix-vector product by neglecting the quadratic non-linearity in the equilibrium distribution function, wherein the vector represents the PDFs, and the matrix represents the collision and streaming operators. Third, we employ classical singular value decomposition (SVD) to decompose the non-unitary collision and streaming operators into a product of unitary matrices. Finally, we show the importance of having a Hadamard gate between the collision and the streaming operations. Our approach has been tested on linear/linearized flow problems such as the advection-diffusion of a Gaussian hill, Poiseuille flow, Couette flow, and lid-driven cavity problems. We provide counts for two-qubit controlled-NOT (CNOT) and single-qubit U gates for test cases involving 9 to 12 qubits, with grid sizes ranging from 24 to 216 points. While the gate count aligns closely with theoretical limits, the high number of two-qubit gates on the order of $10^7$ necessitates careful attention to circuit synthesis.

\end{abstract}


\section{Introduction}

Recent advancements in quantum computing algorithms and the availability of actual quantum processor units (QPUs) offer an exciting opportunity for the field of computational fluid dynamics (CFD) to explore the potential of quantum computing. The CFD workflow generally involves three key steps: mesh generation (for a given geometry), flow modeling (governing equations, discretization, and numerical methods), and post-processing of the results (analysis and visualization). Each of these tasks can demand high computational resources in terms of memory and speed, depending on the problem being solved.
\par
For instance, a three-dimensional direct numerical simulation (DNS) of turbulent flow for a Reynolds number of $10^8$ would require approximately $\Rey^\frac{9}{4} \approx 10^{18}$ grid points \citep{Succi2023,itani2024}. 
Storing a flow variable at each of these $10^{18}$ points using 64-bit precision floating values would require approximately $10^{18} \times 64$ bits of memory, which is equivalent to eight thousand petabytes ($8\times 10^9$ GB) of memory. Quantum superposition and the encoding of classical information on a quantum processor as probability amplitudes suggest the number of qubits required would be approximately 60 ($\sim\frac{15}{2}\log(\Rey)$).  Thus, in terms of computing memory, there is a promising opportunity to perform large-scale CFD simulations on QPUs \citep{Gaitan2020}. However, the QPUs are still in their early stages of development (limited numbers of noisy qubits), and the potential benefits of quantum speed-up have yet to be fully established  \citep{Penuel2024, Sachin2024}. 
\par
Present-day qubits in quantum computers can be affected by errors and imperfections, such as decoherence, gate errors, and measurement errors. In particular, decoherence times are an important factor in determining the performance and feasibility of quantum computations, wherein decoherence refers to the process by which a qubit loses its quantum properties and behaves more classically due to environmental interaction. Thus, the current Noisy Intermediate-Scale Quantum (NISQ) devices are characterized by having a limited number of qubits and relatively short coherence times, which impact their computational capabilities. 

\par
Typical efforts to develop a quantum CFD involve either (i) the application of Schr\"{o}dinger equation to describe fluid motion or (ii) the development of algorithms for solving classical fluid mechanic equations for implementation on QPUs. In the former case, an analogy between the Navier-Stokes equation (NSE) and the Schr\"{o}dinger equation has been derived through Madelung transformation \citep{meng2023}. The latter case focuses on the applicability of existing CFD algorithms portable to QPUs, which received much more attention due to the extensive advancement in CFD algorithms over the past few decades. The main steps to perform CFD on QPUs involve encoding classical data (such as velocity and pressure fields), designing circuits representing differential or integral operators, and employing measurement techniques to convert quantum data back into classical variables. 
\par 
The encoding of classical data requires the preparation of quantum states, which begins with initializing all qubits in their ground state. This is followed by the application of a specific set of quantum gates designed to manipulate the qubits into the desired quantum state. The resulting state vector contains the probability amplitudes corresponding to the classical data.
A key challenge in preparing an arbitrary quantum state for a system comprising $n$ qubits is the exponential increase in the number of required two-qubit controlled-NOT (CNOT) gates. \citet{Sachin2023pnas} has investigated two approaches relevant to CFD that demonstrate reduced CNOT gate requirements, suggesting potential applicability in NISQ devices. In transient flow simulations, reinitialization of the quantum state is necessary at each time step to incorporate updated information, which involves applying a series of quantum gates. This reinitialization process becomes increasingly resource-intensive as both the system size and the number of iterations increase. Moreover, quantum measurements are essential for extracting classical information at each iteration; however, these measurements lead to the collapse of the quantum state, necessitating further reinitialization. This phenomenon introduces additional computational overhead, as multiple measurements may be needed to obtain reliable results, particularly when dealing with complex nonlinear partial differential equations \citep{Schlimgen2022, Succi2023, Sachin2023pnas, Sachin2024}. 
\par
In CFD, the governing equations can be discretized and expressed in matrix-vector form. For explicit time-marching schemes, the discretized equation will be of the form $\boldsymbol{\Phi}^{t+1} = A \boldsymbol{\Phi}^t$, where matrix $A$ represents the discrete differential/integral operator, and vector $\boldsymbol{\Phi}^t$ represents the flow variable at time $t$. In this context, the matrix-vector product $A \boldsymbol{\Phi}^t$ can be viewed as a quantum state evolution, where qubits in a registry are initialized with state vector $\ket{\boldsymbol{\Phi}^t}$ and result in state vector $\ket{\boldsymbol{\Phi}^{t+1}}$ containing the probability amplitudes in the chosen basis state. If the matrix $A$ is not unitary, it can be decomposed as a linear combination of Pauli matrices, or a Hamiltonian $A_H$ can be constructed as
$ A_H = \begin{pmatrix}
  \boldsymbol{0} & iA\\ 
  -i A^T & \boldsymbol{0}
\end{pmatrix} $
which can then be converted to a unitary matrix \cite{childs2012, low2019hamiltonian, brearley2024} as $U(t) = e^{-i A_H t}$.
\par
For implicit schemes, the equation becomes $\boldsymbol{\Phi}^{t+1} = A^{-1} \boldsymbol{\Phi}^t$. In this case, quantum linear solver algorithms such as the Harrow-Hassidim-Lloyd (HHL) algorithm \citep{HHL2009, Cao2012, li2023, Sachin2023pnas, Sachin2024Compact} or the variational quantum linear solver (VQLS) \citep{bravo2023vqls, Pool2024, Ingelmann2024} can be used. VQLS belongs to a class of hybrid algorithms known as variational quantum algorithms (VQAs) that utilize classical optimization schemes on CPUs to provide circuit parameters implementable on NISQ devices. Thus, VQLS performs two tasks simultaneously: quantum circuit parametrization and solution to linear system of equations. In summary, any CFD algorithm can be viewed as a Hamiltonian simulation on QPUs.
\par
Instead of discretizing the NSE, an alternative approach known as the Lattice Boltzmann Method (LBM), which is based on kinetic theory, has been shown to effectively solve the NSE in the incompressible limit. It is important to note that LBM is not restricted to incompressible flows; it can also simulate compressible flows \citep{Alexander1992compress, Xu2012compress, Frapolli2015compress}. LBM describes the evolution of particle distribution functions (PDFs) on a Cartesian grid with a predefined lattice structure. The moments of these PDFs yield the macroscopic flow variables of primary interest, such as density and velocity. Due to particle interactions with their nearest neighbors, LBM demonstrates high parallel efficiency on GPUs.
However, the drawback comes in terms of computational memory. For instance, in an isothermal and incompressible flow simulation with $N$ grid points and $q_n$ lattice directions, storing variables for present and future time steps in LBM will be $2Nq_n$, whereas, in traditional schemes, it would be $2N(D+1)$; $D$-number of physical dimension. The most common lattice structures for 2D and 3D simulations are D2Q9 and D3Q19, which give the scaling factor $q_n = 9$ and $19$, respectively. 
\par
Before the development of the quantum Lattice Boltzmann method (QLBM), \citet{Yepez2001} proposed the quantum lattice-gas model (QLG) for fluid flow simulations that can be implemented using a hybrid quantum-classical nuclear magnetic resonance (NMR) computing methodology, commonly known as Type-II quantum computer (T2QC). Developed in the early 2000s, a T2QC consists of an array of small quantum computers linked by classical communication channels. Several QLG algorithms were developed for T2QC \citep{Yepez2001type2, Berman2002type2, Love2006type2}. Additionally, \citet{PraviaYepez2003} successfully implemented the quantum lattice gas algorithm for the one-dimensional diffusion equation on T2QC. By using two-qubit processors assigned to each of the 16 grid points, they found that the experimental measurements at seven computational time steps aligned well with both the simulation and analytical solutions. This motivated further developments in the QLBM.
\par
Typically, the LBM algorithm is divided into collision and streaming operations. The collision operator is local and has a non-linearity in the velocity, whereas the streaming operator is non-local and linear. Significant work on QLBM can be categorized based on  
(i) tackling the non-linearity \citep{Steijl2022,itani2024, Claudio2024}, 
(ii) circuit design of QLBM \citep{budinski2021ADE, budinski2022NSE}, 
(iii) implementation of streaming without collision \citep{Todorova2020, Schalkers2024JCP}, and
(iv) encoding \citep{Schalkers2024QIP}. 
The fundamental challenge in QLBM involves handling non-linearity in the equilibrium distribution function and its computation through unitaries, which are linear operators \citep{Succi2023}. 
One way to tackle non-linearity is to perform Carleman linearization (CL) of the LB equations, resulting in an infinite dimensional linear system 
\citep{itani2022, li2023, itani2024, Claudio2024, Sanavio2024}. At second-order truncation, the qubit requirement will increase by a factor of two before CL.
Instead of CL which utilize amplitude encoding, \citep{Steijl2022} designed a quantum circuit based on computational basis encoding, which performs quantum floating point arithmetic and thereby evaluate the non-linear term. 
Excluding the non-linear term, \citet{Mezzacapo2015} developed a quantum simulator suitable for trapped ion and superconducting
quantum systems and attempted to solve a two-dimensional advection-diffusion equation (ADE). The collision operator was decomposed as a sum of unitaries and the streaming operator using the D2Q4 lattice was represented as shifted diagonal matrices. Later, \citet{budinski2021ADE} presented a complete circuit with single/ two qubit gates for solving ADE with D1Q3 and D2Q5 lattice model along with linearized equilibrium distribution function (EDF) and constant flow velocity assumptions.
An extension to stream function-vorticity formulation of NSE using QLBM was also attempted with D2Q5 lattice \citep{budinski2022NSE}. In contrast to amplitude encoding, which occupies the whole of Hilbert space, \citet{Schalkers2024QIP} argues that quantum parallelism is best exploited when we utilize the least of Hilbert space. Instead of PDFs, if the velocity were encoded in qubits, neither amplitude encoding nor computational basis encoding technique would permit the collision and streaming to be unitary at once.  
\par 
Based on the above discussion, it is evident that there is no comprehensive quantum model for LBM that encompasses collision, streaming, boundary conditions, and external forcing terms for solving various fluid flow problems in the existing literature. Moreover, treating boundary conditions as a separate operator, dealing with non-orthogonal states, and the possibility of taking measurements at each time step present challenges. As discussed earlier, the measurement-induced collapse of the quantum state further complicates matters, as each measurement alters the state and can potentially introduce errors. Thus, performing simulating without intermediate initialization/measurements is an active field of research in QCFD \citep{Succi2023}.

\par
In this study, we propose the QLB algorithm to solve the 2D low Reynolds number flows with external forcing and wall boundary conditions. Our approach involves representing the streaming and boundary conditions as a unified matrix operator. The objective of this study is to develop QLBM as a Hamiltonian simulation. Our work primarily focuses on implementing the QLBM without making any intermediate assumptions, except for neglecting the non-linear term in the equilibrium distribution function. We note that upon neglecting the non-linearity, the linear attice Boltzmann equation will not recover the incomplete Navier-Stokes equations. However, the linear model is not affected by the nonlinear $O(u^3)$ error terms in the momentum equation \citep{timm_lbm_book}. Thus, the test cases considered herein are linear/linearized flows with low Mach ($\Mach^2 \ll 1 $) and low Reynolds ($\Rey =10$) flows. The quantum circuit presented here for the collsion and streaming remain same throughout the simulation. However, initialization of the quantum state based on the constructed PDFs at the start of each simulation time step is required. 
\par
In the following, the LB model along with the boundary conditions are outlined in Sec.~\ref{sec_lbm}. 
The quantum LB algorithm is described in Sec.~\ref{sec_qlbm}. Results are discussed in Sec.~\ref{sec_results} and Sec.~\ref{sec_summary} summarizes with salient conclusions.


\section{Lattice Boltzmann Method}
\label{sec_lbm}

The  single relaxation time LB model for fluid flow is given by:
\begin{equation}
	f_i(\mathbf{x} + \mathbf{e}_{i} \Delta t, t+\Delta t)  
	= 
	f_i (\mathbf{x},t) - \frac{\Delta t}{\tau}
	\left[ 
	f_i(\mathbf{x},t) - f_i^{eq}(\mathbf{x},t)
	\right]
	    + S_i(\mathbf{x},t)
	\label{eq_lbm_bgk}
\end{equation}
where $f_i$ be the PDF along the $i^{th}$ direction, 
$\mathbf{e}_i$ is the lattice velocity, 
$\tau = 3\nu + 0.5$ is the relaxation parameter with $\nu$ as the kinematic viscosity.
The EDF is given by:
\begin{equation}
	f_i^{eq} = w_{i} \rho \left[ 1+\frac{\mathbf{e}_i\cdot \mathbf{u}}{c_s^2} + \frac{(\mathbf{e}_i\cdot \mathbf{u})^2}{2c_s^4} - \frac{\mathbf{u}\cdot \mathbf{u}}{2c_s^2} \right]
	\label{eq_feq_ns}
\end{equation}
where $w_i$ is the lattice weight, $c_s$ is the sound speed, $\rho$ is the fluid density, and $\mathbf{u}$ is the flow velocity. 
The source term can be evaluated using the simple force model of \citet{Buick2000}, $S_i = w_i c_s^{-2} \mathbf{e}_{i} \cdot \mathbf{F_b}$,
where $\mathbf{F_b}$ is body force. 
Eq.~\eqref{eq_lbm_bgk} is solved using the following two step approach:
\begin{align}
	f_i^*(\mathbf{x}, t) &= 
	f_i (\mathbf{x},t) - \frac{\Delta t}{\tau} \left[ f_i(\mathbf{x},t) - f_i^{eq}(\mathbf{x},t) \right]  + S_i(\mathbf{x},t)
	\label{eq_lb_coll}
\\
f_i(\mathbf{x}, t+\Delta t)  &= f_i^*(\mathbf{x} - \mathbf{e}_{i} \Delta t, t) 
\label{eq_lb_strm}
\end{align}
The Eqs.~\eqref{eq_lb_coll}~and~\eqref{eq_lb_strm} are referred as collision and 
streaming steps, respectively. The moments of distribution function (DF) yields the macroscopic 
flow variables density ($\rho$) and momentum ($\rho\mathbf{u}$),
\begin{align}
		\rho(\mathbf{x}, t) &= \sum_i f_i(\mathbf{x}, t)
		\label{eq_lb_density} \\ 
\rho(\mathbf{x}, t)\mathbf{u}(\mathbf{x}, t) &= \sum_i  \mathbf{e}_i f_i(\mathbf{x}, t)
		\label{eq_lb_momentum}
\end{align}
\noindent
\textit{Boundary Conditions - } 
During streaming step, each lattice site sends and receives the DFs to and from the neighbouring sites. 
In a periodic domain, the DFs leaving on one side of the domain will re-enter on the other side. 
In the case of solid walls, boundary nodes ($\mathbf{x}_b$) send the DFs ($f_i$) to the wall node $x_w$ with 
lattice velocity $\mathbf{e}_i$, which then reflect back with velocity $-\mathbf{e}_i$. 
If the wall is moving with the velocity $u_w$, the reflected DF ($f_{-i}$) will gain/lose the momentum
\citep{timm_lbm_book}. The incoming DF ($f_{-i}$) for the boundary node can be computed using Eq.~\eqref{eq_mov_wall} which includes 
the stationary wall case by taking $u_w=0$. 
\begin{align}
f_{-i}(\mathbf{x}_b, t+\Delta t)  &=  f^*_i(\mathbf{x}, t) - 2 w_{i} \rho_{w} \frac{\mathbf{e}_{i} \cdot \mathbf{u}_{w}}{c_s^2}
\label{eq_mov_wall}
\end{align}


\section{Quantum LB Algorithm}
\label{sec_qlbm}
%

The proposed quantum algorithm consists of three major steps, viz.,
(i) matrix-vector representation of the LB algorithm,
(ii) decomposition of non-unitary LB matrices into unitaries using the singular value decomposition (SVD),
and 
(iii) development of quantum circuit with suitable encoding of PDFs.  In the following, we will explain each of these steps in turn.

\subsection{Matrix-vector representation}
Let $n_g = n_x \times n_y$ be the total number of grid points, where 
$n_x$ and $n_y$ is the number of grid points along $x$ and $y$ axis respectively. 
Then the total number of DFs will be $n_f = n_{\mathbf{e}} n_g$, where $n_{\mathbf{e}}$ is the
number of directions defined in the DmQn lattice model. 
At first, the DFs are arranged in a vector form according to their lattice direction,
$\boldsymbol{df} = [\boldsymbol{df}^1, \boldsymbol{df}^2, \ldots, \boldsymbol{df}^{n_{\boldsymbol{e}}}]$, 
where the superscript refers to the direction. 
The index of any element in $\boldsymbol{df}^{i_e}$ can be written as $x + y n_x + i_e n_x n_y$, 
where $(x,y)$ is the co-ordinates and $i_e$ the direction. 
Next, the entries for the collision and streaming matrices will be derived, 
wherein the row and column indices of the constructed matrix follow the 
same order in which the DFs are arranged. 
In general, every element of the resultant vector from the matrix-vector product can be 
expressed as a linear combination of the elements of the input vector with weights being the 
corresponding row. 
Thus, each $f_i \in \boldsymbol{df}$ can be written as $f_i = \sum\limits_{j=1}^{n_f} \alpha_{ij} f_j$,
 where $\alpha_{ij}$ represents the element at the  $i^{th}$ row and $j^{th}$ column of the matrix.
\par 
\textit{Collision} - 
The quadratic terms in Eq.~\eqref{eq_feq_ns} are needed to recover the NSE through the Chapman-Enskog analysis using the Knudsen number
as the expansion parameter. We assume the flow is low Mach and low Reynolds, and the flow quantities such as pressure, and velocity exhibit 
small variations about the constant rest state. Hence, we neglect the second-order term in Eq.~\eqref{eq_feq_ns}.
Detailed discussion on the validity of linearized EDF can be found in \citet{timm_lbm_book}.

\begin{equation}
	f_i^{eq} = w_{i} \rho \left[ 1+\frac{\mathbf{e}_i\cdot \mathbf{u}}{c_s^2} \right]
	\label{eq_feq_ns_line}
\end{equation}
Using Eqs.~\eqref{eq_lb_density}~and~\eqref{eq_lb_momentum}, 
the linearized EDF in Eq.~\eqref{eq_feq_ns_line} becomes:
\begin{equation} 
f_i^{eq} = \sum\limits_{k=1}^{n_{\mathbf{e}}} w_i (1 + 3 \mathbf{e}_i \cdot \mathbf{e}_k) f_k
\label{eq_feq_ade_comb}
\end{equation}
Now, Eq.~\eqref{eq_lb_coll} can be written as: 
\begin{align} 
f_i^*(\mathbf{x},t) &= (1 - \frac{\Delta t}{\tau}) f_i (\mathbf{x},t) + 
                        \frac{\Delta t}{\tau} f_i^{eq} (\mathbf{x},t) 
            \notag\\
                     &= \sum\limits_{k=1}^{n_{\mathbf{e}}} \Big[ {\delta_{ik}}(1 - \frac{\Delta t}{\tau}) + 
                                    \frac{\Delta t}{\tau} w_i (1 + 3 \mathbf{e}_i \cdot \mathbf{e}_k) 
                                \Big] f_k (\mathbf{x},t)
            \notag\\
                    & = \sum\limits_{k=1}^{n_{\mathbf{e}}} \alpha^{i}_{ik} f_k (\mathbf{x},t)
\label{eq_lb_coll_mod}
\end{align}
where $\delta_{ik}$ is the Kronecker delta function. Note the values of $\alpha^{i}_{ik}$ are constant and
computed once, as they depends on relaxation time, lattice velocity and weights. 
Thus, the matrix form of collision operation (Eq.~\eqref{eq_lb_coll_mod}) can be written as 
%
%
%
\begin{equation}  
  \renewcommand{\arraystretch}{1.75}
 \setcounter{MaxMatrixCols}{13} 
\resizebox{.75\textwidth}{!}{$
\begin{pNiceMatrix}
\RED \alpha_{11}^1 & \cdots & 0 &
\RED \alpha_{12}^2 & \cdots & 0 &
&  &  &
\RED \alpha_{1 n_{\mathbf{e}}}^{n_{\mathbf{e}}} & \cdots & 0
\\
\vdots & \RED  \rdots & \vdots &
\vdots & \RED  \rdots & \vdots &
& \cdots &  &
\vdots & \RED \rdots & \vdots &
\\
0 & \cdots & \RED \alpha_{1 n_g}^1 &
0 & \cdots & \RED \alpha_{1 n_g}^2 &
&  &  &
0 & \cdots & \RED \alpha_{1 n_g}^{n_{\mathbf{e}}} &
\\[7pt]
\RED \alpha_{21}^1 & \cdots & 0 &
\RED \alpha_{22}^2 & \cdots & 0 &
&  &  &
\RED \alpha_{2 n_{\mathbf{e}}}^{n_{\mathbf{e}}} & \cdots & 0
\\
\vdots & \RED \rdots & \vdots &
\vdots & \RED \rdots & \vdots &
& \cdots &  &
\vdots & \RED \rdots & \vdots &
\\
0 & \cdots & \RED \alpha_{2 n_g}^1 &
0 & \cdots & \RED \alpha_{2 n_g}^2 &
&  &  &
0 & \cdots & \RED \alpha_{2 n_g}^{n_{\mathbf{e}}} &
\\[7pt]
 & \vdots & 
 &  & \vdots & 
 & & \cdots &  
 & & \vdots &  
\\
\RED \alpha_{n_{q}1}^1 & \cdots & 0 &
\RED \alpha_{n_{q}2}^2 & \cdots & 0 &
&  &  &
\RED \alpha_{n_{q} n_{\mathbf{e}}}^{n_{\mathbf{e}}} & \cdots & 0
\\
\vdots & \RED \rdots & \vdots &
\vdots & \RED \rdots & \vdots &
& \cdots &  &
\vdots & \RED \rdots & \vdots &
\\
0 & \cdots & \RED  \alpha_{n_{q} n_g}^1 &
0 & \cdots & \RED \alpha_{n_{q} n_g}^2 &
&  &  &
0 & \cdots & \RED \alpha_{n_{q} n_g}^{n_{\mathbf{e}}} &
\end{pNiceMatrix}
\begin{pNiceMatrix}
f^1_1 \\
\vdots\\
f^1_{n_g}\\
f^2_1\\
\vdots\\
f^2_{n_g}\\
\vdots\\ 
f^{n_{\mathbf{e}}}_1\\
\vdots\\
f^{n_{\mathbf{e}}}_{n_g}
\end{pNiceMatrix}
=
\begin{pNiceMatrix}
{f^{1^*}_1} \\
\vdots\\
{f^{1^*}_{n_g}}\\
{f^{2^*}_1} \\
\vdots\\
{f^{2^*}_{n_g}}\\
\vdots\\
{f^{n^*_{\mathbf{e}}}_1} \\
\vdots\\
{f^{n^*_{\mathbf{e}}}_{n_g}}
\end{pNiceMatrix}
$}
\label{mtrx_coll}
\end{equation}
\textit{Streaming and Boundary Conditions} - 
The streaming matrix $S$ consists of $n_{\mathbf{e}} \times n_{\mathbf{e}}$
blocks with the size of each block $n_g \times n_g$. In the absence of boundary
conditions, the streaming operator shift the DFs along the direction of propagation. 
This results in a permutation matrix i.e., 
a square binary matrix where each row/column contains one entry equal to 1,
the remaining entries are zero.
As mentioned earlier, the index of each row $i$ can be decomposed as 
$x + y n_x + i_e n_x n_y$. For a  fully periodic domain, 
the column index $j$ such that $s_{ij}=1$, can be computed by replacing the co-ordinate 
$\mathbf{x}$ by $\mathbf{x} - \mathbf{e}_{i} \Delta t + L$, where $L$ is the domain length. 
In case of a wall boundary, the column index will be computed as 
$x_{\Delta} + y_{\Delta} n_x + i_{-e} n_x n_y$, where 
$(x_{\Delta}, y_{\Delta}) = \mathbf{x} - \mathbf{e}_{i}$ and $i_{-e}$ is the reflection
of the direction $i_e$. 
\par
\begin{figure}[H]
    \centering
\scalebox{0.6}{
    \begin{subfigure}[t]{0.4\columnwidth}
        \includegraphics[width=0.9\textwidth]{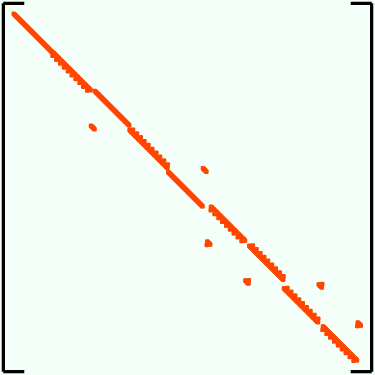}
        \caption{}
        \label{fig_mtrx_ade}
    \end{subfigure}
    \begin{subfigure}[t]{0.4\columnwidth}
        \includegraphics[width=0.9\textwidth]{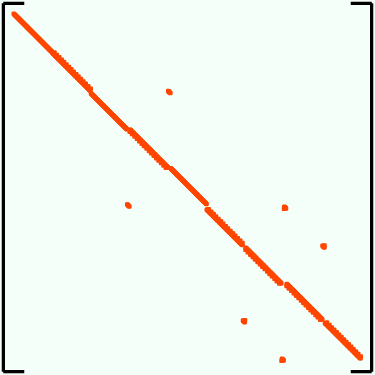}
        \caption{}
        \label{fig_mtrx_pois}
    \end{subfigure}
    \begin{subfigure}[t]{0.4\columnwidth}
        \includegraphics[width=0.9\textwidth]{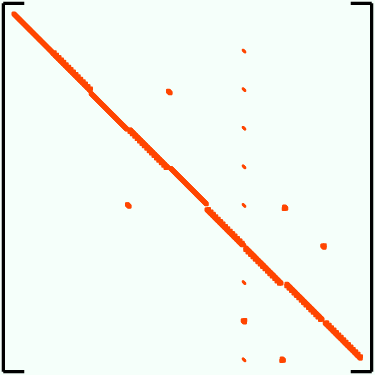}
        \caption{}
        \label{fig_mtrx_couet}
    \end{subfigure}
    \begin{subfigure}[t]{0.4\columnwidth}
        \includegraphics[width=0.9\textwidth]{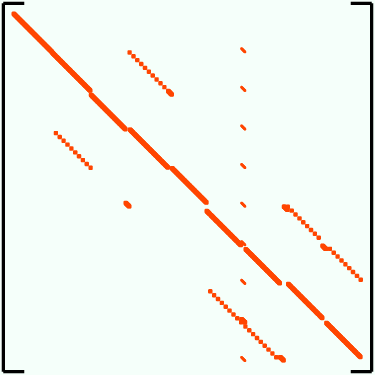}
        \caption{}
        \label{fig_mtrx_lid}
    \end{subfigure}
}
\caption{Streaming matrices for test cases with $n_{qa}=11$ qubits using D2Q9 lattice model (a) a fully periodic domain,
(b) periodic along $x$ axis and bounded by stationary walls in $y$ axis
(c) periodic along $x$ axis and bounded by stationary (bottom) and moving walls (top) in $y$ axis
(d) bounded by a moving top wall and stationary walls on other three sides.}
\label{fig_mtrx_stream}
\end{figure}

 
As an illustration, we choose the D2Q9 lattice model, which is a commonly used lattice for 
two dimensional LB fluid flow simulations \citep{timm_lbm_book}. The lattice parameters 
are given in Table.~\ref{tab_d2q9}. 
The first block of matrix $S$ will be identity, as the lattice velocity will be zero.
In a fully periodic case, the number of elements in the streaming matrix will be the size of diagonal entries and 
the row major ordering of grid points provide the shifted elements closer to the diagonal (Fig.~\ref{fig_mtrx_ade}). 
The matrix shapes in Fig.~\ref{fig_mtrx_stream} are obtained for the test cases presented 
in Sec.~\ref{sec_results} with $n_{qa}=11$ and the grid sizes are given in Table.~\ref{tab_case_gates}.
%
\begin{table}[H]
\centering
\renewcommand{\arraystretch}{1.2}%
\caption{Lattice parameters for D2Q9 model.}
\begin{tabular*}{0.95\columnwidth}{c @{\extracolsep{\fill}} c c } 
\thickhline
$i$ & $\mathbf{e}_{i}$ & $w_i$ \\
\hline
0 & $(0,0)$ & 4/9 \\
1,2,3,4 & $(0,1), (1,0), (0,-1),(-1,0)$ & 1/9 \\
5,6,7,8 & $(1,1), (1,-1), (-1,1),(-1,-1)$ & 1/36 \\
\thickhline
\end{tabular*}
\label{tab_d2q9}
\end{table}
%
\subsection{Unitary Decomposition}
Neither the collision nor the streaming matrix generated in the previous section is unitary. There are several methods to decompose non-unitary matrices into unitary matrices. One common approach is to express them as a weighted sum of unitaries, known as the Linear Combination of Unitaries  (LCU) method \citep{childs2012}, or as a product of unitaries, such as through Singular Value Decomposition (SVD) \citep{gilyen2019}. The number of terms required in an LCU representation of a matrix generally depends on three factors: (i) the size of the matrix, (ii) the sparsity or structure of the matrix, and (iii) the precision and algorithmic context of the LCU application. For a general $N \times N$ matrix, the worst-case scenario typically requires $O(N)$ terms. But, for sparse matrices, the number of terms can be proportional to the number of non-zero entries, resulting in $O(k)$, where $k$ is the number of non-zero terms.
In contrast to LCU, the number of matrices produced in SVD is always three, with two of them being unitary. 
However, \citet{Sachin2024} utilized four-unitaries approach, wherein a non-unitary matrix can be approximated as a sum of four unitaries using an expansion parameter. Thus, finding the efficient decomposition of a given non-unitary matrix is highly crucial and is a challenging area of research \citep{Berry2015, An2023LCUoptimal, Sachin2024}.
\par
In the present work, we use the singular value decomposition (SVD) to transform collision/streaming matrices as a combination of unitary matrices. Specifically, a matrix $A$ can be decomposed into $A = U*D*V$, where $U,V$ are unitary and $D$ a diagonal matrix but need not be an unitary. Again, this presents two approaches for representing D: either as a sum or as a product of unitaries. Previosuly, \citet{Schlimgen2022} introduced a dilation-based algorithm based on SVD (product of unitaries) to simulate a non-unitary diagonal operator. In our work, we utilized LCU to decompose the diagonal matrix as a sum of two unitary matrices. Before we decompose $D$, the norm of the elements in $D$ may not be unity. Hence, we  normalize it and store the factor $\alpha$ for the post multiplication of resultant vector. Now, $D$ can be decomposed as $D = 0.5\ (D_1+D_2)$, where $D_1, D_2$  are unitary matrices given by $D_1, D_2 = D \pm i\ \sqrt{I-D^2}$. Thus, $A$ will be written in terms of unitaries $U *  {\frac{1}{2} (D_1+D_2) } * V$. Consequently, the total number of unitary operators in our quantum LB algorithm is eight. In contrast, in the generic LCU approach, this count depends on the size of the matrix, which scales with the problem size. 

\subsection{Quantum Algorithm}
The algorithm consists of three steps: encoding, circuit construction, and measurement. 
Herein, we encode the PDFs as the sole variable using the amplitude encoding technique.
The collision and streaming operations are performed via the quantum circuits built from 
the  unitaries. Finally, we measure and evaluate the probability amplitudes resulting in the 
PDFs for the next time step.  The distinct advantage of the present algorithm is that 
the circuit is built once for the entire simulation. 
This means the circuit built at the start of the simulation can be iterated using a $\textit{for}$ loop,
with the statevector re-initialized at each time step of the simulation.
The complete circuit model for implementing QLBM is given in Fig.~\ref{fig_qlb_circuit}.
\begin{figure*}
\begin{adjustbox}{width=0.95\textwidth, center}
    \input{circuit}
\end{adjustbox}
\caption{Quantum circuit for LBM, wherein the collision and streaming blocks are repeated for $n$ time steps.}
\label{fig_qlb_circuit}
\end{figure*}

\par 
\textit{Encoding - }
Two quantum registers, a computational ($q$) and an ancilla ($a$) will be used. 
The $q$ register contains $n_q = \log_2(n_{\mathbf{e}} \times n_x \times n_y)$ qubits,
whereas ancilla register has one qubit. In case $2^{n_q} > n_f$, additional 
zeros are padded to make the vector $\boldsymbol{df}$ of length $2^{n_q}$.
Because of ancilla qubit, the size of statevector $\ket{\phi}$ is of $2^{1+n_q}$.
Now, the quesion is whether to control the initialization on a state of the ancilla i.e., 
$\boldsymbol{df}$ is either to append with itself or to pad with zero entries.
Physically, this means whether the ancilla qubit should ($(H \otimes I)( \ket{0}_a \otimes \ket{\phi} )$) or 
should not ($\ket{0}_a \otimes  \ket{\phi}$) be in superposition
with computational qubits. Because of non-essential orthogonal states generated after collision, the answer is affirmative,
and it will be discussed in the Appendix.
We append the vector $\boldsymbol{df}$ to itself, resulting in a vector of length $2^{1+n_q}$.
Thus, initial state of qubits $\phi = [\boldsymbol{df}, \boldsymbol{df}]$ will be encoded as,
\begin{equation}
\ket{\boldsymbol{\phi}^0} = \sum_{i=1}^{2^{1+n_q}} \frac{\phi_i}{\lVert \phi \rVert} \ket{i}
\end{equation}
\par 
\textit{Quantum circuits - }
At first, the collision matrices are converted into quantum gates by the direct application 
of \textit{UnitaryGate} operation, which is available in major quantum software development kits (SDKs), such as 
Qiskit. After SVD decomposition, the collision matrix can be written as 
$C = U^c *{\frac{1}{2} (D^c_1+D^c_2) } * {V^c}$. 
Because of the summation, matrices $D^c_1$ and $D^c_2$ cannot be directly applied as gates.
Following \citep{budinski2021ADE, low2019hamiltonian}, the controlled operation for the 
linear combination of unitaries will be implemented as,
\begin{equation}
(H^\dagger \otimes I_q)(\ket{0}\bra{0}_a) \otimes D^c_1 + \ket{1}\bra{1}_a) \otimes D^c_2)(H \otimes I_q)
\label{eq_circ_diag_unit}
\end{equation}
The unitary corresponding to Eq.~\eqref{eq_circ_diag_unit} is 
$\begin{pmatrix}
  D^c & i\sqrt{I -D^c {D^c}^\dagger}\\ 
  i\sqrt{I-{D^c}{D^c}^\dagger} & D^c
\end{pmatrix}$. The single ancilla is used as a control qubit, which through 
Hadamard gate ($H$) is in superposition with the computational qubits.
Thus the complete collision operation as product of unitaries can be expressed as
\begin{equation}
\begin{pmatrix}
  U^c & \boldsymbol{0} \\ 
  \boldsymbol{0}  & U^c
\end{pmatrix}
\begin{pmatrix}
  D^c & i\sqrt{I -D^c {D^c}^\dagger}\\ 
  i\sqrt{I-{D^c}{D^c}^\dagger} & D^c
\end{pmatrix}
\begin{pmatrix}
  {V^c} & \boldsymbol{0} \\ 
  \boldsymbol{0}  & {V^c}
\end{pmatrix}
\label{eq_coll_unit_mat}
\end{equation}
Upon collision, the quantum state $\ket{\boldsymbol{\phi}^0}$ is transformed to $\ket{\boldsymbol{\phi}^c}$ as
\begin{equation}
\ket{\boldsymbol{\phi}^c} = \frac{1}{{\lVert \phi \rVert} \cdot \lVert  \alpha_c \rVert}  (\ket{0}_a \sum_{i=1}^{2^{n_q}} \phi^c_i \ket{i}_q + \ket{1_\phi}_{aq})
\end{equation}
where $ \lVert  \alpha_c \rVert$ denotes the normalization factor obtained from the elements of diagonal matrix $D^c$. 
Similarly, the streaming operation can be expressed as a product of unitaries,
\begin{equation}
\begin{pmatrix}
  U^s & \boldsymbol{0} \\ 
  \boldsymbol{0}  & U^s
\end{pmatrix}
\begin{pmatrix}
  D^s & i\sqrt{I -D^s {D^s}^\dagger}\\ 
  i\sqrt{I-{D^s}{D^s}^\dagger} & D^s
\end{pmatrix}
\begin{pmatrix}
  {V^s} & \boldsymbol{0} \\ 
  \boldsymbol{0}  & {V^s}
\end{pmatrix}
\label{eq_strm_unit_mat}
\end{equation}
\par
However, before proceeding to streaming after collision, the Hadamard needs to be applied 
on the ancilla i.e. circuit $(H \otimes I_q)$ has to be applied. In the Appendix we show its necessity.
Upon performing streaming the real part in the first block of statevector results in the desired PDF for the next simulation time step \cite{bautista2021}.
\par 
\textit{Measurement \& flow variables - }
The measurement process in quantum computing does not yield the complete 
state vector of qubits in the register. Instead it results in the collapse of the so-called
quantum wave function. Here, we encoded PDFs onto the amplitude of the standard basis.
Thus, the measurement collapses to one of the state $\ket{j}$ and contains the binary 
string of zeros and ones. Since a quantum device is a probabilistic model, we 
execute the circuit multiple times and compute the probability distribution of 
the different states of the qubit. This results in the distribution function of LBM.
Now, converting the DFs into the flow variable can be done in the post-processing 
step on the classical computer. The matrix form for evaluating the 
first and second moments of PDFs 
in Eqs.~\eqref{eq_lb_density}~and~\eqref{eq_lb_momentum} is given by 
\begin{eqnarray}
M^0 \text{:   } 
\begin{pmatrix}
  I^1 & \cdots &  I^{n_{\mathbf{e}}}  
\end{pmatrix}
\begin{pmatrix}
\boldsymbol{df}^1 \\
\vdots\\
\boldsymbol{df}^{n_{\mathbf{e}}}  
\end{pmatrix}
& = 
\begin{pmatrix}
\boldsymbol{\rho}
\end{pmatrix}
\label{eq_mtrx_m0}
\\
M^1 \text{:   }  
\begin{pmatrix}
  e^1_x I & \cdots & e^{n_{\mathbf{e}}}_x I  \\
  e^1_y I & \cdots & e^{n_{\mathbf{e}}}_y I
\end{pmatrix}
\begin{pNiceMatrix}
\boldsymbol{df}^1 \\
\vdots\\
\boldsymbol{df}^{n_{\mathbf{e}}}  
\end{pNiceMatrix}
& = 
\begin{pNiceMatrix}
\boldsymbol{\rho} \mathbf{u_x} \\
\boldsymbol{\rho} \mathbf{u_y}
\end{pNiceMatrix}
\label{eq_mtrx_m1}
\end{eqnarray}
\noindent
where the identity matrices $I$ in Eqs.~\eqref{eq_mtrx_m0}~and~\eqref{eq_mtrx_m1}
are of size $n_g \times n_g$ and repeated $n_{\mathbf{e}}$ times horizontally.
Upon computing variables such as density and momentum, the normalization factor $({\lVert \phi \rVert} \cdot {\lVert  \alpha_c \rVert}\cdot {\lVert  \alpha_s \rVert})$ need to multiplied; where ${\lVert  \alpha_s \rVert}$ is the norm of elements in $D^s$.


\section{Results and Discussion}
\label{sec_results}

The proposed algorithm is implemented using IBM's Qiskit quantum computing SDK.
Since we are interested in obtaining PDFs, we used the state vector simulator in 
Qiskit-Aer package \citep{Qiskit}. The unitaries derived in Eq.~\eqref{eq_coll_unit_mat}~and~\eqref{eq_strm_unit_mat} are applied 
in the Qiskit-simulator as a custom gate operation. Four test cases with different boundary condintions were chosen for study.
The D2Q9 lattice is used unless otherwise mentioned.
%

\subsection{Advection-Diffusion of a Gaussian Hill}
\begin{figure}[H]
\begin{adjustbox}{width=0.6\columnwidth, center}
\includegraphics[scale=0.37]{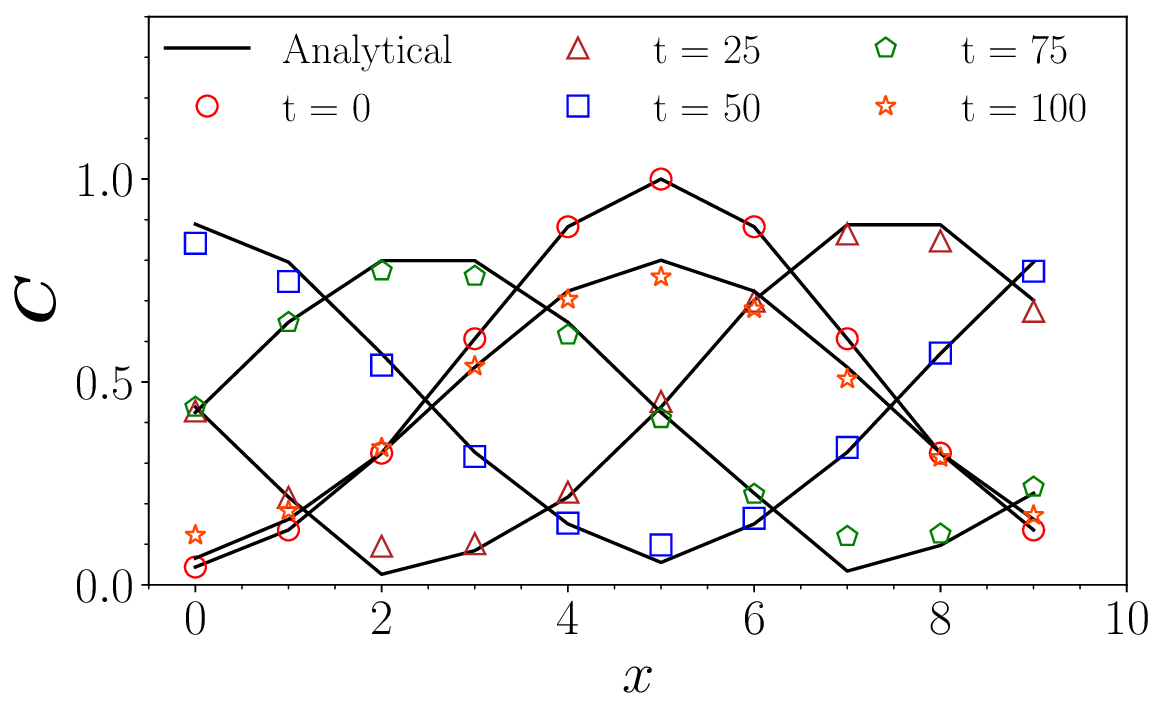}
\end{adjustbox}
\caption{Comparison of concentration profiles obtained from QLBM with analytical solution (\citet{timm_lbm_book}) for the advection-diffusion of Gaussian hill of grid size $10 \times 10$ at different time steps. The parameters are: relaxation time $\tau=0.515$, $\boldsymbol{C}_0=1.0$, $\sigma_0=2.0$, and $D = 0.005$.}
\label{fig_ade_compare}
\end{figure}

\begin {figure*}[b!]
\begin{adjustbox}{width=0.95\textwidth, center}
\includegraphics[width=0.99\textwidth]{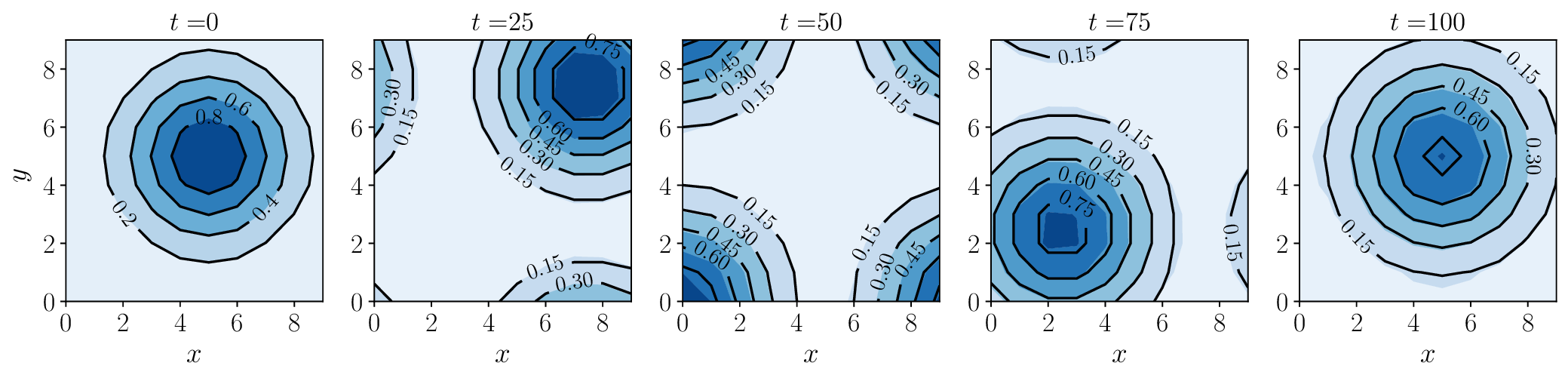}
\end{adjustbox}
\caption{Comparison of concentration contours obtained from QLBM with analytical solution (bold lines) (\citet{timm_lbm_book}) for the advection-diffusion of Gaussian hill of grid size $10 \times 10$. The chosen parameters are specified in caption of Fig.~\ref{fig_ade_compare}.}
\label{fig_ade_contour}
\end{figure*}

The evolution of species concentration $\boldsymbol{C}$ with an initial Gaussian 
profile is simulated in a fully periodic two-dimensional domain.
The time evolution of concentration field $\boldsymbol{C}$ in the presence of
a homogeneous advection velocity $\mathbf{u}$ is given by an analytical solution \citep{timm_lbm_book},
\begin{equation}
\boldsymbol{C}(\mathbf{x},t) = \frac{\sigma^2_0}{\sigma^2_0 + \sigma^2_D} \boldsymbol{C}_0 \exp{\frac{-(\mathbf{x-x_0-u}t)^2}{2\sigma^2_0+\sigma^2_D}}
\end{equation}
where $\boldsymbol{C}_0$ the maximum concentration value, 
$\sigma_0$ is the initial mean squared deviation and $\sigma_D=\sqrt{2Dt}$, $D$ is diffusion coefficient. 
Initially,
\begin{equation}
\boldsymbol{C}(\mathbf{x},t=0) = \boldsymbol{C}_0 \exp{\frac{-(\mathbf{x-x_0})^2}{2\sigma^2_0}}
\end{equation}
where $\mathbf{x_0}$ is chosen at the center of the domain.
We choose the domain of grid size $10\times 10$, which results in $n_f=900$ and $n_q = 10$. 
The values of other parameters are: $\boldsymbol{C}_0=1.0$, $\sigma_0=2.0$, and $D = 0.005$.
The relaxation time can be computed by $\tau=3D+0.5$. 
The simulation ran for 100 time steps. Fig.~\ref{fig_ade_compare} shows the comparison between the analytical solution and QLBM results 
of $\boldsymbol{C}$ profiles picked along the $x$-axis where the $\max\{\boldsymbol{C}\}$ is found.
The comparison of concentration contours between QLBM results with analytical solution is given in Fig.~\ref{fig_ade_contour}.
After $100$ time steps, the $\mathrm{L}^2$ relative error norm ($=\sqrt{\frac{\sum (\boldsymbol{C}_{\text{ref}} - \boldsymbol{C}_{\text{LBM}})^2}{\sum \boldsymbol{C}^2_{\text{ref}}}}$) is found to be around $6\%$.

\subsection{Plane Poiseuille Flow}
\begin{figure}
\begin{center}
\includegraphics[scale=0.55]{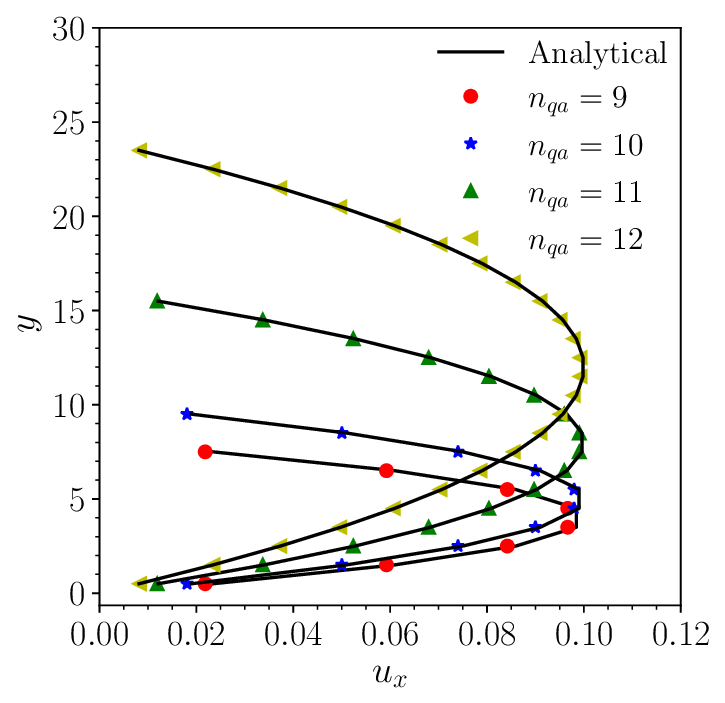}
\end{center}
\caption{The horizontal velocity profile of plane Poiseuille flow  obtained from QLBM for four different qubit counts is compared with analytical solution (\citet{timm_lbm_book}). The values of $F_b$ are 0.001, 0.008, 0.0005 and 0.0003, respectively for the $n_{qa}=$ 9, 10, 11 and 12. The grid size and the final time step are given in Table.~\ref{tab_pois_l2}.}
\label{fig_pois_comp}
\end{figure}

%
The plane Poiseuille flow is an important test case of pressure driven flow,
where a constant pressure gradient is applied in the direction of flow.
The domain is two-dimensional which is periodic along the flow direction ($x$-axis) 
and bounded by stationary walls at top and bottom  ($y$-axis). The analytical 
solution for horizontal velocity distribution along the vertical axis ($y$) is given by \citep{timm_lbm_book},
\begin{equation}
u_x(y) = \frac{G}{2\mu} y(y-h)
\end{equation}
where $G$ is the pressure gradient applied along the direction of the flow,
$\mu$ is the dynamic viscosity, and 
$h$ is the distance between the walls. 
The Reynolds number ($\Rey$) of the flow is taken as $10$. The value for the forcing term in $S_i$ can be 
evaluated as $\mathbf{F_b} = \frac{8\nu}{h^2}\mathbf{u}_{max}$, with $\mathbf{u}_{max}=(0.1,0)$.
Four different grid sizes were chosen by varying the number of qubits $n_{qa}$ (Table.~\ref{tab_pois_l2}).
Fig.~\ref{fig_pois_comp} shows the comparison of horizontal velocity profile along the $y$-axis taken at the 
center of $x-axis$. The maximum relative error of around $2\%$ is found in the smaller grid case of size $3\times 8$ which utilized $9$ qubits (Table.~\ref{tab_pois_l2}).
\begin{table}[H] 
\caption{Relative error norm obtained for Poiseuille flow test case with different grid sizes.}
\renewcommand{\arraystretch}{1.15}%
\begin{tabular*}{0.95\columnwidth}{c @{\extracolsep{\fill}} c c l}
\thickhline
\multirow{2}{*}{$n_{qa}$} & \multirow{2}{*}{Grid} & \multirow{2}{*}{\# Timesteps} & \multirow{2}{*}{$\mathrm{L}^2$ (\%)} \\
 & & &  \\
 \hline
9  &   $3 \times 8$  & 500 & 2.34  \\
10 &   $5 \times 10$ & 700 & 1.35  \\
11 &   $7 \times 16$ & 900 & 0.6   \\
12 &   $9 \times 24$ & 1200 & 0.56 \\
\thickhline
\end{tabular*}
\label{tab_pois_l2}
\end{table}


\subsection{Plane Couette-Poiseuille Flow}
\begin {figure}
\begin{adjustbox}{width=0.45\columnwidth, center}
\includegraphics[scale=0.4]{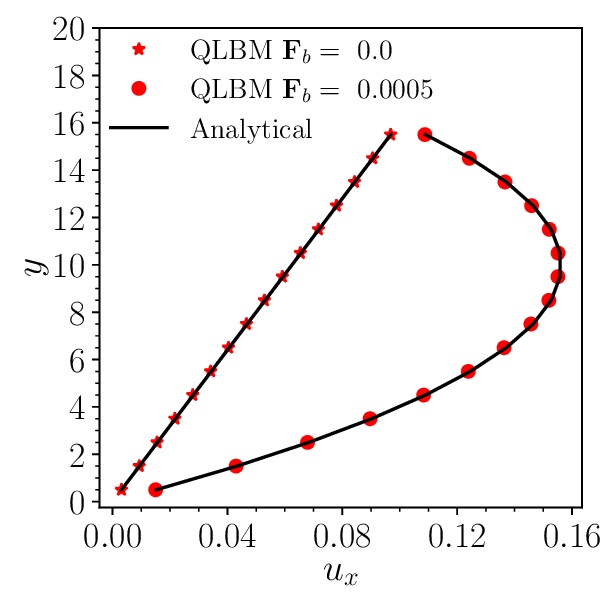}
\end{adjustbox}
\caption{The horizontal velocity profile of plane Couette flow obtained from QLBM for grid size of $7 \times 16$ with $n_{qa}=11$ is compared with analytical solution (\citet{timm_lbm_book}) at the end of time step $t=900$.}
\label{fig_coute_comp}
\end{figure}

The plane Couette-Poiseuille flow is a test case of shear induced fluid motion. 
Test case set-up is similar to the Poiseuille flow except that it features a moving wall at the top boundary.
The analytical solution for the horizontal velocity distribution along the vertical axis ($y$) is given by
\begin{equation}
u_x(y) = u_w \frac{y}{h} + \frac{G}{2\mu} y(y-h)
\end{equation}
Here the top wall velocity is taken as $u_w=0.1$, and all other parameters are chosen as in Poiseuille flow case.
The grid size of $7 \times 16$ with $n_{qa}=11$ is chosen. Two test cases with and without pressure gradient has been simulated. The simulation ran for $900$ time steps, and compared against the analytical solution (Fig.~\ref{fig_coute_comp}). The relative difference is found to be $0.3\%$.

\subsection{Lid Driven Cavity}
\begin{figure}
\centering
\begin{adjustbox}{width=0.5\columnwidth, center}
\includegraphics[scale=0.55]{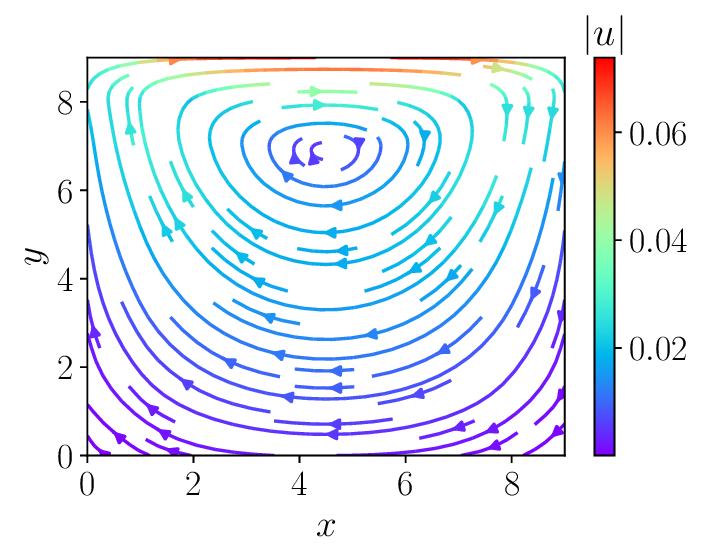}
\end{adjustbox}
\caption{Lid-driven cavity case in $10 \times 10$ grid. Streamlines for $Re=10$ obtained from QLBM at the end of time step $t=1000$.}
\label{fig_lid_streamline}
\end{figure}

\begin{figure}
\centering
\begin{adjustbox}{width=0.45\columnwidth, center}
\includegraphics[scale=0.5]{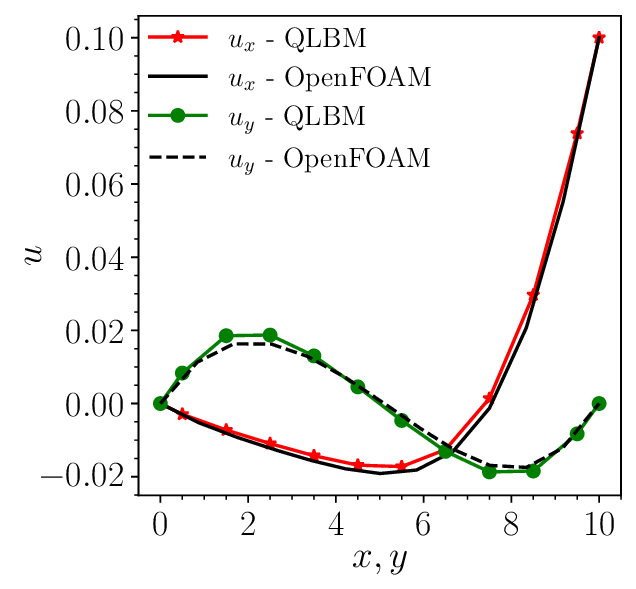}
\end{adjustbox}
\caption{Lid-driven cavity case $Re=10$ in $10 \times 10$ grid. Comparison of $x$ and $y$ velocities from QLBM with OpenFOAM solution. The $x$ and $y$ velocities are obtained along the $y$ and $x$-axis respectively.}
\label{fig_lid_compare}
\end{figure}

The lid-driven cavity is a standard CFD benchmark problem for validating various numerical schemes for low-speed flows. A square cavity with two side walls and the bottom are considered to be rigid, whereas the top wall is moving with a tangential velocity $u_w$. Before proceeding with QLBM, we performed the classical LB simulation with and without non-linear terms in the equilibrium distribution function. We choose the grid size of $10 \times 10$. The lid velocity is chosen as $u_w=0.1$ and $Re=10$. This gives $\nu=u_wL/Re=0.1$, and $\tau=3\nu+0.5=0.8$. This setup yielded no significant difference between the LB simulations with and without non-linearity at low Re. However, we observed a significant difference when $Re=100$, necessitating a larger grid size to maintain stability. Hence, in order to validate the QLBM simulation with the coarse grid in the absence of an analytical solution, we used \textit{icoFOAM} solver from OpenFOAM-v2312 to obtain the reference solution \citep{jasak2009openfoam}. Simulation time of $1 \si{s}$ with $\Delta t = 0.005 \si{s}$ is used in OpenFOAM, whereas QLBM simulation is performed for $1000$ time steps. Fig.~\ref{fig_lid_streamline} shows the streamlines obtained from QLBM  after $1000$ time steps. Fig.~\ref{fig_lid_compare} shows the comparison of QLBM results with OpenFOAM for horizontal ($u_x$) and vertical ($u_y$) velocity profiles extracted, respectively along the $y$ and $x$-axis at the center of the domain. The relative difference is found to be $5.9\%$.  
%
%
\subsection{Quantum Gates and Computational Complexity}
%

\begin{table*}[!t]
\caption{Gate count obtained from Poiseuille flow test case solved for different grid sizes (single time step).}
\begin{adjustbox}{width=0.99\textwidth, center}
\renewcommand{\arraystretch}{1.15}%
\begin{tabular}{c @{\extracolsep{\fill}} c c r r r r r r c @{\extracolsep{\fill}} r r r r r r} 
\thickhline
\multirow{2}{*}{$n_{qa}$} & & \multirow{2}{*}{Grid points} & \multicolumn{6}{c}{CNOT} & & \multicolumn{6}{c}{U} \\ 
\cline{4-9}\cline{11-16}
 &   & & $U^c$ & $D^c$ &$V^c$ &$U^s$ &$D^s$ & $V^s$ & & $U^c$ & $D^c$ &$V^c$ &$U^s$ &$D^s$ & $V^s$  \\
\hline 
9  &  & 24 $(3 \times 8)$ & 31020 & 5325 & 31020 & 31020 & 5325 & 16505 & & 62168 & 4100 & 62168 & 62160 & 4069 & 31435
\\
10 &  & 50 $(5 \times 10)$ & 124844 & 12362 & 124844 & 124844 & 12362 & 64959 & & 249942 & 9854 & 249944 & 249775 & 9599 & 126602
\\
11 &  & 112 $(7 \times 16)$ & 500908 & 27619 & 500908 & 500908 & 27619 & 256491 & & 1002326 & 20901 & 1002327 & 1001847 & 20838 & 504379
\\
12 &  & 216 $(9 \times 24)$ & 2006700 & 61307 & 2006700 & 2006700 & 61307 & 1016860 & & 4014422 & 45816 & 4014424 & 4013528 & 45561 & 2014951
\\
\thickhline
\end{tabular}
\end{adjustbox}
\label{tab_pois_gates}
\end{table*}

\begin{table*}[!t]
\caption{Gate count obtained for all test cases with $n_{qa}=11$ (single time step).}
\begin{adjustbox}{width=0.99\textwidth, center}
\renewcommand{\arraystretch}{1.15}%
\begin{tabular}{l @{\extracolsep{\fill}} c r r r r r r c @{\extracolsep{\fill}} r r r r r r} 
\thickhline
\multirow{2}{*}{Case} & \multirow{2}{*}{Grid points} & \multicolumn{6}{c}{CNOT} & & \multicolumn{6}{c}{U} \\ 
\cline{3-8}\cline{10-15}
 &  & $U^c$ & $D^c$ &$V^c$ &$U^s$ &$D^s$ & $V^s$ & & $U^c$ & $D^c$ &$V^c$ &$U^s$ &$D^s$ & $V^s$  \\
\hline 
ADE & 100 $(10 \times 10)$ & \multirow{4}{*}{500908} & \multirow{4}{*}{27619} & \multirow{4}{*}{500908} & \multirow{4}{*}{500908} & \multirow{4}{*}{27619} & 134397 & & 1002327 & 21093 & 1002325 & 1002066 & 20838 & 257167
\\
Poiseuille & 112 $(7 \times 16)$  &  &  &  &  &  & 256491 & & 1002326 & 20901 & 1002327 & 1001847 & 20838 & 504379
\\
Couette & 112 $(7 \times 16)$     &  &  &  &  &  & 500908 &  & 1002324 & 20901 & 1002327 & 1002328 & 21861 & 1002327
\\
Cavity & 100 $(10 \times 10)$ &  &  &  &  &  & 500908 &  & 1002326 & 21093 & 1002328 & 1002328 & 21861 & 1002322
\\
\thickhline
\end{tabular}
\end{adjustbox}
\label{tab_case_gates}
\end{table*}

Implementing the present algorithm on real QPU hardware is challenging for several reasons. 
Firstly, when applying Hadamard gates to all qubits including the ancilla, all possible states in the Hilbert space have equal probabilities. Since we encode PDFs onto these probabilities, we must prepare the state of qubits before the start of QLBM. The number of CNOT gates required for state preparation  \citep{shende2005} will be $O(2^n)$. In the present algorithm, at each time step, LBM is expressed as a product of six unitary matrices, three each for collision and streaming operation. Of the six unitaries, two are diagonal matrices.
Each of these matrix decomposed as controlled-NOT (CNOT) and U-gates using Qiskit decomposition tool, which uses the 
quantum Shannon decomposition (QSD) algorithm. In QSD, the circuit for $n$-qubit operator is decomposed into three rotational gates 
and $n-1$ qubits operators; and this undergoes recursively until the circuit is made of single and two qubit gates \citep{shende2005}.
Table.~\ref{tab_pois_gates} shows the gate count obtained from the Poiseuille flow case with four different qubit counts. 
For each additional qubit, gate counts of both CNOT and U gate increases by a factor of $4$, which matches with the theoretical estimate of $O(4^{n_{qa}})$.
Each diagonal operator over $n_{qa}$ qubits will be decomposed into $O(2^{n_{qa}+1})$ gates, and the general unitary
operator require $O(2^{n_{qa}-1} (2^{n_{qa}} -1))$ gates. 
Thus, the total gate count for two diagonal matrices and four generic unitaries is of $O(2^{n_{qa}+2} (2^{n_{qa}} - 1))$.
Next, we report on the influence of different streaming-cum-boundary operators on the gate count.
Table.~\ref{tab_case_gates} shows the gate count obtained for all test cases with $n_{qa}=11$.
The difference is observed in the streaming part $V^s$ where the gate count of fully bounded case (Lid-cavity)
is four times that of fully periodic case (ADE). 
\par
Another challenging aspect of the quantum simulation described by generic unitary matrices is the system memory of classical computers. The quantum state described by the probability amplitudes will be stored as an array of complex numbers. 
A single qubit system described by two complex numbers requires two 64-bit precision floating values i.e., $2\times2\times64=256$ bits $=32$ bytes of memory. An arbirary $2\times 2$ unitary matrix representing a quantum operation requires $4\times2\times64=512$ bits $=64$ bytes of memory. In case of LBM with D2Q9 lattice with $n_{qa} = 20$ qubits, which can handle a grid size of $1024 \times 1024$, storing statevector and the single unitary matrix would require $0.017$, and $18000$
Gigabytes (GB) of memory, respectively. Because our algorithm comprises of six unitaries, simulation with larger qubits are not feasible, unless
the matrix can be built directly from the universal quantum gate set. It is worth to mention the importance of non-linear term on the computational requirements. For instance, Carleman system at second order truncation \citep{Claudio2024} requires $n_{qa} = \log_2(n_f + n_f^2 )$ , 
whereas the present work requires $n_{qa} = \log_2(n_f)$. Also, the computational cost of SVD on the classical computer is of 
$O(2^{3n_{qa}})$, which we've not taken into account.

\section{Summary and Outlook}
\label{sec_summary}
%

We presented a comprehensive quantum model for LBM that encompasses collision, streaming, boundary conditions, and external forcing terms for solving various fluid flow problems in the low $\Rey$ regime of $O(10)$. As a result of neglecting the non-linear term in the equilibrium function, the LB algorithm is written in terms of a matrix-vector product. Further, we combined the streaming operation and boundary conditions as a single matrix operator. We utilized classical SVD to decompose the non-unitaries, which the followed by representing the diagonal operator as a sum of unitaries. Thus, we have eight unitary operators describing the LB algorithm for a single time step. Compared to the generic linear combination of unitaries approach, where the number of unitaries depends on the problem size, the fixed number of unitaries is a key advantage of the present algorithm. The algorithm has been tested with four different CFD benchmark problems performed using the 'statevector' simulator. The results are in good agreement with the reference solutions. 
\par
The present QLB algorithm has certain limitations. First, the quantum state evolution performed with any of eight unitary matrices generated in a single time step of QLBM results in a larger circuit depth of $O(10^5)$ (minimum estimate among all test cases), which is not feasible in present day QPUs, as they are made of qubits with shorter decoherence times and their two qubit gates are noisy/error-prone. But the quantum circuit presented here for the collsion and streaming remain same throughout the simulation. This allows us to explore various optimization strategies to obtain the shallow circuit depth, which in turn has a potential to be implementable on NISQ devices. However, initialization of the quantum state based on the constructed PDFs at the start of each simulation time step is required. Second, simulations with more than 15 qubits using the D2Q9 model are not feasible due to higher memory requirements. Even though the first limitation is hardware-related, circuit depth can be optimized by building a parameterized quantum circuit or performing a unitary circuit synthesis with more efficient algorithms. This helps efficiently implement the Carleman system for LBM, which includes non-linear terms, as it requires at least twice the resources required by the present algorithm. Further works include testing the present algorithm in real hardware or simulation with noise models. This helps to quantify (i) the single qubit noise on the final distribution functions and (ii) the affect of number of measurements required to construct the state vector.

\section*{Acknowledgments}
This research was kindly supported by the Quantum Computing Consortium that has been funded by the MAGNET program of the Israel Innovation Authority. 




\bibliographystyle{apalike}
\bibliography{references}
\section*{Appendix}
Let's denote the matrices defined in Eq.~\eqref{eq_coll_unit_mat} and \eqref{eq_strm_unit_mat} as,
\begin{align}
U_c &= \begin{pmatrix}
  U^c & \boldsymbol{0} \\ 
  \boldsymbol{0}  & U^c
\end{pmatrix} \\
D_c &= \begin{pmatrix}
  D^c & i\sqrt{I -D^c {D^c}^\dagger}\\ 
  i\sqrt{I-{D^c}{D^c}^\dagger} & D^c
\end{pmatrix}
\\
V_c &= \begin{pmatrix}
  {V^c} & \boldsymbol{0} \\ 
  \boldsymbol{0}  & {V^c}
\end{pmatrix}
\\
U_s &= \begin{pmatrix}
  U^s & \boldsymbol{0} \\ 
  \boldsymbol{0}  & U^s
\end{pmatrix}
\\
D_s &= \begin{pmatrix}
  D^s & i\sqrt{I -D^s {D^s}^\dagger}\\ 
  i\sqrt{I-{D^s}{D^s}^\dagger} & D^s
\end{pmatrix}
\\
V_s &= \begin{pmatrix}
  {V^s} & \boldsymbol{0} \\ 
  \boldsymbol{0}  & {V^s}
\end{pmatrix}
\end{align}
The complete LB operation can be written as (operation from right to left) as 
\begin{equation}
\ket{\phi^1} = U_s D_s V_s U_c D_c V_c \ket{\phi^0}
\label{eq_app_lbm}
\end{equation}
where $\ket{\phi^0}$ and $\ket{\phi^1}$ are the statevectors containing initial and the final DFs, respectively. 
As mentioned earlier, the choice of encoding $\boldsymbol{df}$ with padded zeros of same size
found to be unsuitable. Let $\ket{\phi^0} = \begin{pmatrix}  \boldsymbol{df} \\ \boldsymbol{0} \end{pmatrix}$, where $\boldsymbol{0}$
is the zero vector of size same as $\boldsymbol{df}$. We'll apply the matrix operations 
defined in Eq.~\eqref{eq_app_lbm} (from right to left). Thus,
\begin{align}
V_c \ket{\phi^0} &= \begin{pmatrix}  V^c \boldsymbol{df} \\ \boldsymbol{0} \end{pmatrix}
\\
D_c V_c \ket{\phi^0} &= \begin{pmatrix}  D^c V^c \boldsymbol{df} \\  i\sqrt{I -D^c {D^c}^\dagger} V^c \boldsymbol{df} \end{pmatrix}
\\
U_c D_c V_c \ket{\phi^0} &= \begin{pmatrix}  U^c D^c V^c \boldsymbol{df} \\  i U^c \sqrt{I -D^c {D^c}^\dagger} V^c \boldsymbol{df} \end{pmatrix}
\label{eq_encode_1_issue}
\\
V_s U_c D_c V_c \ket{\phi^0} &= \begin{pmatrix}  V^s U^c D^c V^c \boldsymbol{df} \\  i V^s U^c \sqrt{I -D^c {D^c}^\dagger} V^c \boldsymbol{df} \end{pmatrix}
\\
%
D_s V_s U_c D_c V_c \ket{\phi^0} &= \begin{pmatrix}  D^s V^s U^c D^c V^c \boldsymbol{df} - \sqrt{I -D^s {D^s}^\dagger} V^s U^c \sqrt{I -D^c {D^c}^\dagger} V^c \boldsymbol{df}     \\ i \sqrt{I-{D^s}{D^s}^\dagger} V^s U^c D^c V^c \boldsymbol{df} + i D^s V^s U^c \sqrt{I -D^c {D^c}^\dagger} V^c \boldsymbol{df}  \end{pmatrix}
\\
U_s D_s V_s U_c D_c V_c \ket{\phi^0} &= \begin{pmatrix} U^s D^s V^s U^c D^c V^c \boldsymbol{df} - U^s \sqrt{I -D^s {D^s}^\dagger} V^s U^c \sqrt{I -D^c {D^c}^\dagger} V^c \boldsymbol{df}     \\ i U^s \sqrt{I-{D^s}{D^s}^\dagger} V^s U^c D^c V^c \boldsymbol{df} + i U^s D^s V^s U^c \sqrt{I -D^c {D^c}^\dagger} V^c \boldsymbol{df} \end{pmatrix}
\label{eq_app_df_0} 
\end{align}
Thus, the imaginarity present in the second block of Eq.~\eqref{eq_encode_1_issue} results in the additional term \\
$U^s \sqrt{I -D^s {D^s}^\dagger} V^s U^c \sqrt{I -D^c {D^c}^\dagger} V^c \boldsymbol{df} $ in Eq.~\eqref{eq_app_df_0} affecting the post-streaming state vector.
Now, let's choose $\ket{\phi^0} = \begin{pmatrix}  \boldsymbol{df} \\ \boldsymbol{df} \end{pmatrix}$, and apply the matrix operations as before,
\begin{align}
V_c \ket{\phi^0} &= \begin{pmatrix}  V^c \boldsymbol{df} \\ V^c \boldsymbol{df} \end{pmatrix}
\\
D_c V_c \ket{\phi^0} &= \begin{pmatrix}  (D^c + i\sqrt{I -D^c {D^c}^\dagger}) V^c \boldsymbol{df} \\  (D^c + i\sqrt{I -D^c {D^c}^\dagger}) V^c \boldsymbol{df} \end{pmatrix}
\\
U_c D_c V_c \ket{\phi^0} &= \begin{pmatrix}  U^c (D^c + i\sqrt{I -D^c {D^c}^\dagger}) V^c \boldsymbol{df} \\ U^c (D^c + i\sqrt{I -D^c {D^c}^\dagger}) V^c \boldsymbol{df} \end{pmatrix} \label{eq_app_coll_hd}
\\
V_s U_c D_c V_c \ket{\phi^0} &= \begin{pmatrix}  V^s U^c (D^c + i\sqrt{I -D^c {D^c}^\dagger}) V^c \boldsymbol{df} \\ V^s U^c (D^c + i\sqrt{I -D^c {D^c}^\dagger}) V^c \boldsymbol{df} \end{pmatrix}
\\
D_s V_s U_c D_c V_c \ket{\phi^0} &= \begin{pmatrix} (D^s + i\sqrt{I -D^s {D^s}^\dagger}) V^s U^c (D^c + i\sqrt{I -D^c {D^c}^\dagger}) V^c \boldsymbol{df} \\ (D^s + i\sqrt{I -D^s {D^s}^\dagger}) V^s U^c (D^c + i\sqrt{I -D^c {D^c}^\dagger}) V^c \boldsymbol{df} \end{pmatrix}
\\
U_s D_s V_s U_c D_c V_c \ket{\phi^0} &= \begin{pmatrix} U^s (D^s + i\sqrt{I -D^s {D^s}^\dagger}) V^s U^c (D^c + i\sqrt{I -D^c {D^c}^\dagger}) V^c \boldsymbol{df} \\ U^s (D^s + i\sqrt{I -D^s {D^s}^\dagger}) V^s U^c (D^c + i\sqrt{I -D^c {D^c}^\dagger}) V^c \boldsymbol{df} \end{pmatrix}
\\
&=
\begin{pmatrix} 
\hspace{-4em}U^s D^s V^s U^c D^c V^c \boldsymbol{df} - U^s \sqrt{I -D^s {D^s}^\dagger} V^s U^c \sqrt{I -D^c {D^c}^\dagger} V^c \boldsymbol{df} \\
\hspace{4em} + i( U^s D^s V^s U^c \sqrt{I -D^c {D^c}^\dagger} V^c \boldsymbol{df} + U^s \sqrt{I -D^s {D^s}^\dagger} V^s U^c D^c V^c \boldsymbol{df})
\\
\hspace{-4em}U^s D^s V^s U^c D^c V^c \boldsymbol{df} - U^s \sqrt{I -D^s {D^s}^\dagger} V^s U^c \sqrt{I -D^c {D^c}^\dagger} V^c \boldsymbol{df} \\
\hspace{4em} + i( U^s D^s V^s U^c \sqrt{I -D^c {D^c}^\dagger} V^c \boldsymbol{df} + U^s \sqrt{I -D^s {D^s}^\dagger} V^s U^c D^c V^c \boldsymbol{df})
\end{pmatrix}
\label{eq_encode_2_issue}
\end{align}
Thus, the final statevctor in Eq.~\eqref{eq_encode_2_issue} contains undesired terms in both real and imaginary components.
In order to get the correct DFs, before proceeding to streaming, the Hadamard needs to be applied on the ancilla i.e. circuit $(H \otimes I_q)$ has to be applied. The resulting matrix operation is defined as $H^{cs} = (H \otimes I_q) = \frac{1}{\sqrt{2}} 
\begin{pmatrix}
  I & I \\ 
  I  & -I
\end{pmatrix}$. After collision operation defined in Eq.~\eqref{eq_app_coll_hd}, we proceed as follows:
\begin{align}
H^{cs} U_c D_c V_c \ket{\phi^0} &= \frac{1}{\sqrt{2}} \begin{pmatrix} 2(U^c (D^c + i\sqrt{I -D^c {D^c}^\dagger}) V^c \boldsymbol{df}) \\ \boldsymbol{0} \end{pmatrix}
\\
V_s H^{cs} U_c D_c V_c \ket{\phi^0} &= \frac{1}{\sqrt{2}} \begin{pmatrix} 2(V^s U^c (D^c + i\sqrt{I -D^c {D^c}^\dagger}) V^c \boldsymbol{df}) \\ \boldsymbol{0} \end{pmatrix}
\\
D_s V_s H^{cs} U_c D_c V_c \ket{\phi^0} &= \frac{1}{\sqrt{2}} \begin{pmatrix} 2( D^s V^s U^c (D^c + i\sqrt{I -D^c {D^c}^\dagger}) V^c \boldsymbol{df}) \\ 2( i\sqrt{I -D^s {D^s}^\dagger} V^s U^c (D^c + i\sqrt{I -D^c {D^c}^\dagger}) V^c \boldsymbol{df}) \end{pmatrix}
\\
U_s D_s V_s H^{cs} U_c D_c V_c \ket{\phi^0} &= \frac{1}{\sqrt{2}} \begin{pmatrix} 2( U^s D^s V^s U^c (D^c + i\sqrt{I -D^c {D^c}^\dagger}) V^c \boldsymbol{df}) \\ 2( U^s i\sqrt{I -D^s {D^s}^\dagger} V^s U^c (D^c + i\sqrt{I -D^c {D^c}^\dagger}) V^c \boldsymbol{df}) \end{pmatrix}
\\
&= \sqrt{2} \begin{pmatrix} U^s D^s V^s U^c D^c V^c \boldsymbol{df} + i U^s D^s V^s U^c \sqrt{I -D^c {D^c}^\dagger} V^c \boldsymbol{df} \\  iU^s\sqrt{I -D^s {D^s}^\dagger} V^s U^c (D^c + i\sqrt{I -D^c {D^c}^\dagger}) V^c \boldsymbol{df} \end{pmatrix}
\end{align}
Finally, the real part in the first block of statevector results in the desired PDF for the next simulation time step.

\end{document}